\documentclass{article}

\usepackage{arxiv}

\usepackage[utf8]{inputenc} % allow utf-8 input
\usepackage[T1]{fontenc}    % use 8-bit T1 fonts
\usepackage{hyperref}       % hyperlinks
\usepackage{url}            % simple URL typesetting
\usepackage{booktabs}       % professional-quality tables
\usepackage{amsfonts}       % blackboard math symbols
\usepackage{nicefrac}       % compact symbols for 1/2, etc.
\usepackage{microtype}      % microtypography
\usepackage{lipsum}		    % Can be removed after putting your text content
\usepackage{graphicx}
\usepackage{natbib}
\usepackage{titlesec}
\usepackage{doi}

\graphicspath{ {./images/} }

\title{\textit{Precision-medicine-toolbox}: An open-source python package for facilitation of quantitative medical imaging and radiomics analysis}

\author{ \href{https://orcid.org/0000-0002-3856-9740}{\includegraphics[scale=0.06]{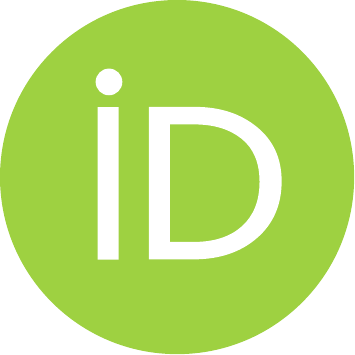}\hspace{1mm}Sergey Primakov} \\
	The D-Lab, Department of Precision Medicine\\
	GROW—School for Oncology, Maastricht University\\
	6200 MD Maastricht, The Netherlands \\
	\texttt{s.primakov@maastrichtuniversity} \\
	\And
	\href{https://orcid.org/0000-0003-2751-790X}{\includegraphics[scale=0.06]{orcid.pdf}\hspace{1mm}Elizaveta Lavrova} \\
	The D-Lab, Department of Precision Medicine\\
	GROW—School for Oncology, Maastricht University\\
	6200 MD Maastricht, The Netherlands \\
	GIGA Cyclotrone Research Center\\
	University of Liege, Belgium\\
	\texttt{e.lavrova@maastrichtuniversity.nl} \\
	\And
	\href{https://orcid.org/0000-0002-9900-329X}{\includegraphics[scale=0.06]{orcid.pdf}\hspace{1mm}Zohaib Salahuddin} \\
	The D-Lab, Department of Precision Medicine\\
	GROW—School for Oncology, Maastricht University\\
	6200 MD Maastricht, The Netherlands \\
	\texttt{z.salahuddin@maastrichtuniversity.nl} \\
	\And
	\href{https://orcid.org/0000-0001-7911-5123}{\includegraphics[scale=0.06]{orcid.pdf}\hspace{1mm}Henry Woodruff} \\
	The D-Lab, Department of Precision Medicine\\
	GROW—School for Oncology, Maastricht University\\
	6200 MD Maastricht, The Netherlands \\
	Department of Radiology and Nuclear Medicine \\
	Maastricht University Medical Centre \\
	Maastricht, The Netherlands \\
	\texttt{h.woodruff@maastrichtuniversity.nl} \\
	\And
	\href{https://orcid.org/0000-0001-7961-0191}{\includegraphics[scale=0.06]{orcid.pdf}\hspace{1mm}Philippe Lambin} \\
	The D-Lab, Department of Precision Medicine\\
	GROW—School for Oncology, Maastricht University\\
	6200 MD Maastricht, The Netherlands \\
	Department of Radiology and Nuclear Medicine \\
	Maastricht University Medical Centre \\
	Maastricht, The Netherlands \\
	\texttt{philippe.lambin@maastrichtuniversity.nl} \\
}

\renewcommand{\shorttitle}{\textit{arXiv} Template}

\hypersetup{
pdftitle={Precision-medicine-toolbox: An open-source python package for facilitation of quantitative medical imaging and radiomics analysis},
pdfsubject={precision-medicine-toolbox},
pdfauthor={Sergey Primakov, Elizaveta Lavrova, Zohaib Salahuddin, Henry Woodruff, Philippe Lambin},
pdfkeywords={Medical imaging research, DICOM, Radiomics, Statistical analysis, Features, Pre-processing},
}
\begin{document}
\maketitle

\begin{abstract}
Medical image analysis plays a key role in precision medicine as it allows the clinicians to identify anatomical abnormalities and it is routinely used in clinical assessment. Data curation and pre-processing of medical images are critical steps in the quantitative medical image analysis that can have a significant impact on the resulting model performance.
In this paper, we introduce a \textit{precision-medicine-toolbox} that allows researchers to perform data curation, image pre-processing and handcrafted radiomics extraction (via Pyradiomics) and feature exploration tasks with Python. With this open-source solution, we aim to address the data preparation and exploration problem, bridge the gap between the currently existing packages, and improve the reproducibility of quantitative medical imaging research.
\end{abstract}

% keywords can be removed
\keywords{Medical imaging research \and DICOM \and Radiomics \and Statistical analysis \and Features \and Pre-processing}

\includegraphics[scale=0.24]{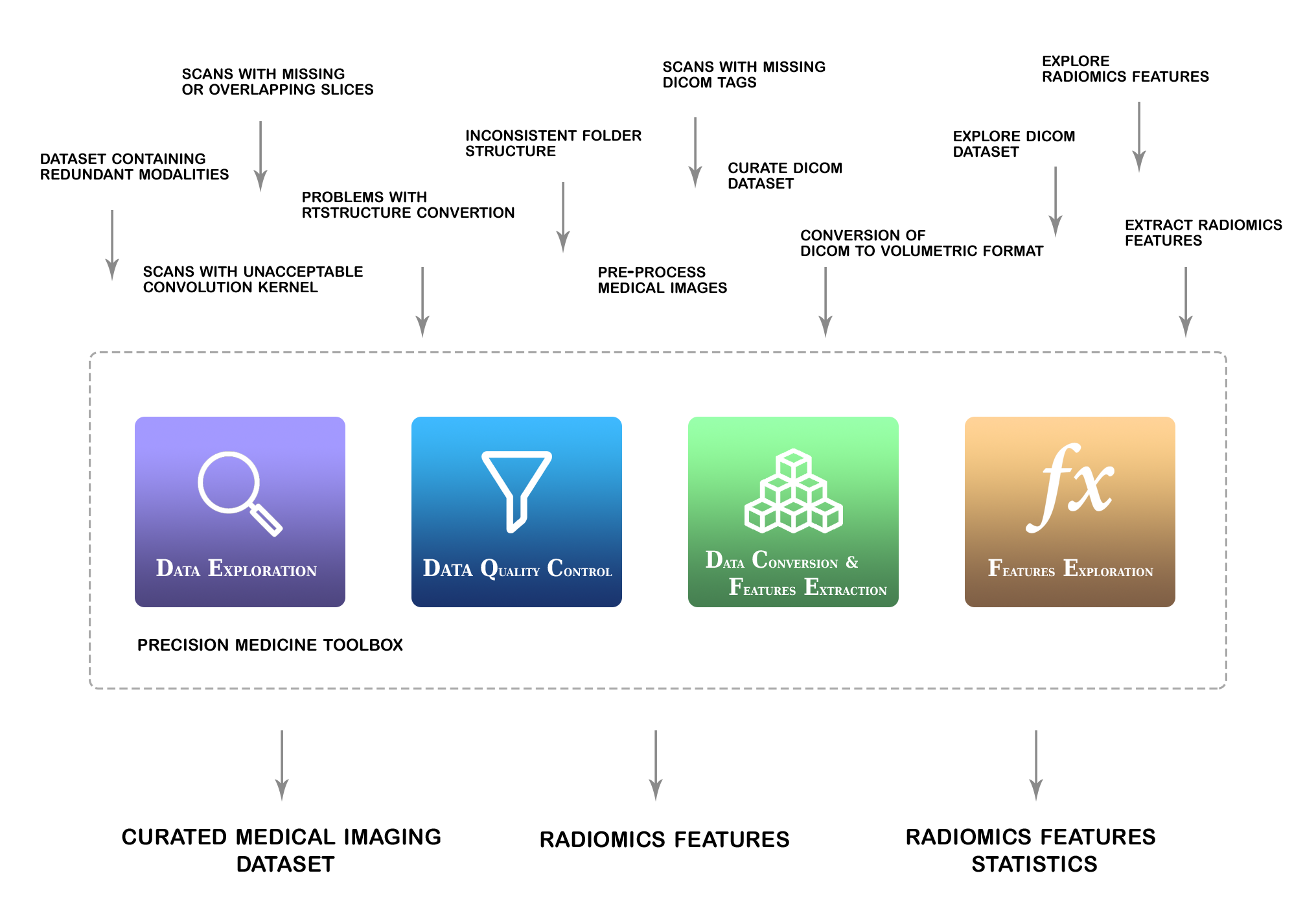}

\section{Introduction}
Medical imaging allows the visualisation of anatomical structures and metabolic processes of the human body and plays an integral part in clinical decision-making for diagnostic, prognostic, and treatment purposes (\cite{Beheshti2021-gr,Wei2019-kk}). Medical imaging is becoming increasingly popular in clinical practice due to increasing accessibility of hardware, rising population and growing confidence in the utility of multiple imaging modalities (\cite{Smith-Bindman2008-yk}). Precision medicine aims to enhance individual patient care by identifying subgroups of patients within a disease group using genotypic and phenotypic data for better understanding of the disease characteristics and consequently targeting the disease with more precise treatment (\cite{Niu2019-gw,Carrier-Vallieres2018-ml}). Medical image analysis plays a key role in precision medicine as it allows the clinicians to identify anatomical abnormalities and it is routinely used in clinical assessment (\cite{Acharya2018-zi}). 

The amount of healthcare imaging data from disparate imaging sources is exploding and it is not possible for radiologists to cope up with the increasing demand. Multiple studies have shown that there is a significant inter-observer variability for various clinical tasks (\cite{Kinkel2000-dd,Luijnenburg2010-ls}). Hence, there is a need for quantitative image analysis tools to aid the clinicians in meeting the challenges of rising demand and better clinical performance. Radiomics is the extraction of quantitative image features and correlating them with biological and clinical outcomes (\cite{Lambin2017-kg}). The field of radiomics is gaining traction each year due to increase in computational power and increasing amount of multimodal data (\cite{Oren2020-eb,Aggarwal2021-cm,Zhou2021-ky}) as illustrated by Figure 1. The field of radiomics has demonstrated promising results in various clinical applications including diagnostics, prognosis and decision support systems (\cite{Tagliafico2020-jp,Zhang2017-so,Wang2021-zx,Mu2020-pw}). Radiomics can broadly be classified into two different categories: handcrafted radiomics and deep learning. Handcrafted radiomics utilises machine learning techniques and image biomarker standardisation initiative (IBSI)-compliant handcrafted features (such as shape, intensity, and texture features) extracted from a specific region of interest (\cite{Rogers2020-hx}). Pyradiomics is one of the available open source tools that allows the extraction of IBSI-compliant handcrafted radiomics features (\cite{Van_Griethuysen2017-ph}). Deep learning automatically learns representative image features from the high dimensional image data without the need of feature engineering by using non-linear modules that constitute a neural network (\cite{Schmidhuber2015-la}). Convolutional neural networks are deep neural networks that became popular in 2012 after the AlexNet architecture demonstrated state-of-the-art performance for image recognition (\cite{Krizhevsky2017-ms}). Since then, convolutional neural networks have demonstrated state-of-the-art performance for many clinical tasks (\cite{Murtaza2020-vm,Bhatt2021-dj,Mazurowski2019-kp}). Tensorflow (\cite{Abadi2016-oa}) with Keras interface (\cite{Gulli2017-xi}) and Pytorch (\cite{Paszke2019-my}) are popular deep learning frameworks for the implementation of deep neural networks. 
\break\break\break
\includegraphics[scale=0.8]{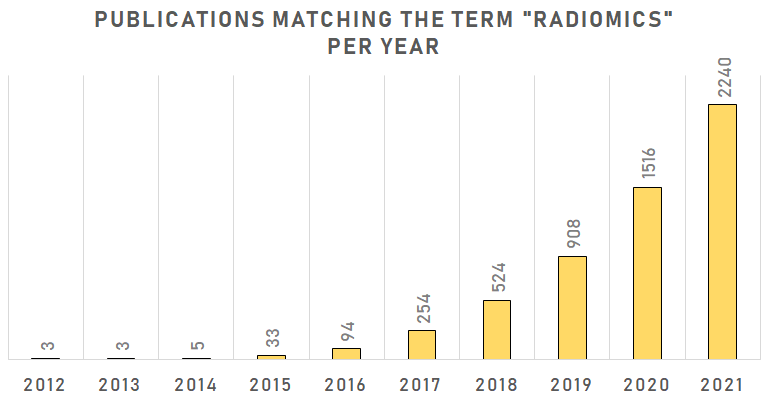}
\begin{tabular}{c}
Figure 1 -  Number of publications, by year, containing the keyword ‘radiomics’ in PubMed database\\
(https://pubmed.ncbi.nlm.nih.gov/?term=radiomics\&timeline=expanded).
\end{tabular}
\break\break
Data Curation and pre-processing of medical images are time-taking and critical steps in the radiomics workflow that can have a significant impact on the resulting model performance (\cite{Fave2016-ux,Hosseini2021-bm,Zhang2019-ha}). The data curation step usually comprises several steps such as image format conversion, out-of-distribution detection and checks for redundant modalities, unacceptable convolution kernel, and missing or overlapping slices. These steps may be performed manually or using low level python libraries such as Pydicom (\cite{Mason2011-kt}), Nibabel (\cite{Brett2020-bm}), SimpleITK (\cite{Yaniv2018-nh}), Numpy (\cite{Van_der_Walt2011-mn}), Pandas (\cite{McKinney2011-rb}), Scipy (\cite{Virtanen2020-jy}), Scikit-image (\cite{Van_der_Walt2014-ih}), and Scikit-learn (\cite{Kramer2016-zr}). The re-implementation of the above-mentioned data curation steps by the researchers makes it error-prone and results in increased difficulty for reproducibility. It is also important to investigate the potential of image processing during the development of a radiomics workflow. Image biomarker standardisation initiative (IBSI) also emphasises on the need of image processing before the extraction of radiomics features (\cite{Zwanenburg2020-ca}). Moreover, it is also important to perform an exploratory analysis on the handcrafted radiomics features and visualise discriminatory statistics. While there are available tools for the implementation of entire radiomics pipeline such as Nipype (\cite{Gorgolewski2016-vf}), Pymia (\cite{Jungo2021-fg}), and MONAI (\cite{MONAI_Consortium2020-kr}), there is still a need of a toolbox that allows researchers to perform critical tasks such as data curation, image pre-processing and handcrafted radiomics feature exploration during the development of the radiomics study. 

In this paper, we introduce the \textit{precision-medicine-toolbox} that allows researchers to perform data curation, image pre-processing and handcrafted radiomics feature exploration tasks. This toolbox will also benefit the researchers without a strong programming background to implement these critical steps and increase the reproducibility of quantitative medical imaging research. In this paper, we discuss the functionality of the first release of this open source project. In future, more functionality will be added to the toolbox.

\section{Methods}

\subsection{Example data}
The functionality of the toolbox is demonstrated on the Lung1 open-source dataset. The Lung1 dataset (\cite{Jungo2021-fg,MONAI_Consortium2020-kr,Aerts2014-jv}) contains pretreatment CT scans of 422 non-small cell lung cancer (NSCLC) patients, as well as manually delineated gross tumor volume (GTV) for each patient, and clinical outcomes. The imaging data is presented in Digital Imaging and Communications in Medicine (DICOM) format. The delineations are available in Radiotherapy Structure (RT Structure) format. The clinical data is present in Comma-Separated Values (CSV) format.

\subsection{Design and implementation}
\subsubsection{Organisation of the toolbox}
The toolbox allows for the preparation of the imaging datasets and exploration of the feature datasets. As illustrated in Figure 2, dedicated base classes have been implemented for each dataset type (imaging or features) to extract the corresponding data, as well as the associated metadata. The functionality classes inherit from the base classes. This approach allows for the separation of reading and processing tasks and makes it readily available for new data formats or functions.
\break\break\break
\includegraphics[scale=0.38]{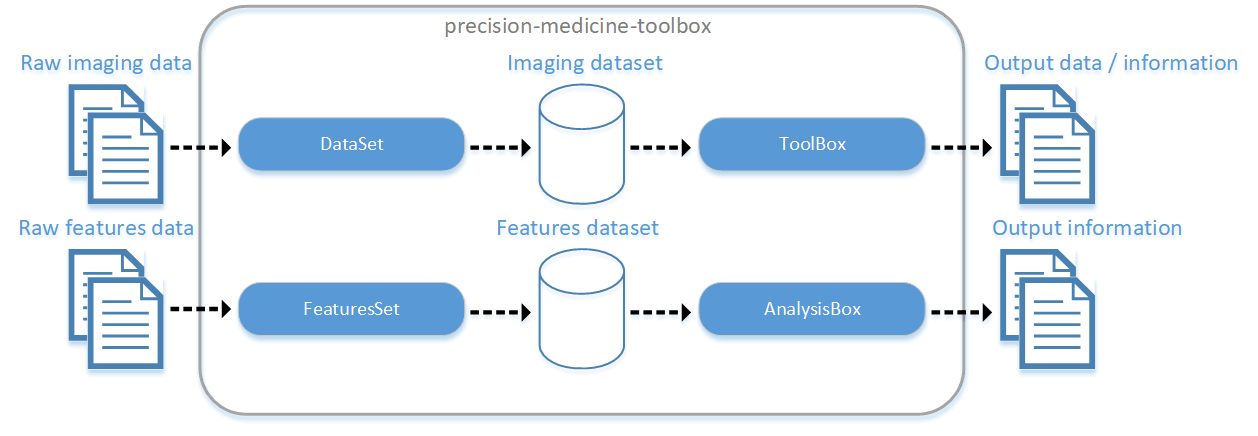}
\begin{tabular}{c}
Figure 2 - Organisation of the \textit{precision-medicine-toolbox}: The \textit{DataSet} class takes an imaging dataset as in input \\and is inherited by the \textit{ToolBox} class; the \textit{FeaturesSet} class takes a features dataset as an input and is inherited by\\ the \textit{AnalysisBox} class. 
\end{tabular}
\break

\subsubsection{Imaging module}

This module allows for pre-processing and exploration of the imaging datasets. It consists of the base \textit{DataSet} class and the inheriting \textit{ToolBox} class. The \textit{DataSet} class reads the imaging data and the corresponding metadata and initializes a dataset object. The \textit{ToolBox} class allows for high-level functionality while working with the raw computed tomography (CT) or magnetic resonance (MR) imaging data. Currently, the following functions are implemented:
\begin{itemize}
    \item dataset parameters exploration by parsing of the imaging metadata,
    \item dataset basic quality examination, including check of imaging modality, slice thickness, number of slices, in-plane resolution and pixel spacing, and reconstruction kernel,
    \item conversion of DICOM dataset into volumetric Nearly Raw Rusted Data (NRRD) dataset,
    \item image basic pre-processing, including bias field correction, intensity rescaling and normalization, histogram matching, intensities resampling, histogram equalization, image reshaping,
    \item unrolling NRRD images and ROI masks into Joint Photographic experts Group (JPEG) slices for a quick check of the converted images or any existing NRRD or MetaImage Medical Format (MHA) dataset,
    \item radiomics  features extraction from NRRD/MHA data using PyRadiomics package (\cite{Van_Griethuysen2017-ph}).
\end{itemize}

\subsubsection{Features module}
This module allows for the exploration of the feature datasets. It consists of the base \textit{FeaturesSet} class and the inheriting \textit{AnalysisBox} class. The \textit{FeaturesSet} class reads the features data and the corresponding metadata, and initializes a \textit{FeaturesSet} object. The \textit{AnalysisBox} class allows for the basic analysis of the features. Currently, the following functions are implemented:
\begin{itemize}
    \item visualization of feature values distributions in classes,
    \item visualization of features mutual Spearman correlation matrix,
    \item calculation of corrected p-values for Mann-Whitney test for features mean values in groups,
    \item visualization of univariate receiver operating characteristic (ROC) curves for each feature and calculation of the area under the curve (AUC),
    \item volumetric analysis, including visualization of volume-based precision-recall curve and calculation of Spearman correlation coefficient between every feature and volume,
    \item calculation of basic statistics (number of missing values, mean, std, min, max, Mann-Whitney test p-values for binary classes, univariate ROC AUC for binary classes, Spearman's correlation with volume if volumetric feature name is sent to function) for every feature.
\end{itemize}

\subsection{Online documentation and tutorials}
The online documentation for the \textit{precision-medicine-toolbox} contains information about the source code, third-party packages, package installation, quick start, instructions for contribution, information about the authors, code licence, and acknowledgments. The examples of the toolbox functionality implementation are presented in tutorials. The full description of the classes and methods is presented in Application Programming Interface (API) specifications.

\section{Results}

\subsection{Design and implementation}

\subsubsection{Organisation of the toolbox}
The \textit{precision-medicine-toolbox} is implemented in Python (Python Software Foundation, Wilmington, DA, U.S.) and requires version 3.6 or higher. The source code is hosted on GitHub (https://github.com/primakov/precision-medicine-toolbox) and Zenodo platform (DOI 10.5281/zenodo.6126913). It depends on the following packages: NumPy (\cite{Harris2020-eu}), SimpleITK (\cite{Lowekamp2013-lh}), Tqdm (\cite{Lowekamp2013-lh,Da_Costa-Luis2019-ab}), Pydicom (\cite{Mason2011-kt}), Pandas (\cite{Mason2011-kt,McKinney2010-me}), PyRadiomics (\cite{Van_Griethuysen2017-ph}), Scikit-image (\cite{Van_der_Walt2014-ih}), Ipywidgets (\cite{Jupyter-widgets_undated-et}), Matplotlib (\cite{Hunter2007-su}), Pillow (\cite{Clark2015-pb}), Scikit-learn (\cite{Buitinck2013-nf}), Scipy (\cite{Buitinck2013-nf,Virtanen2020-jy}), Plotly (\cite{noauthor_undated-wl}), Statmodels (\cite{Seabold2010-xr}). The \textit{precision-medicine-toolbox} package has been released under the BSD-3-Clause License and is available from the Python Package Index (PyPI) repository (https://pypi.org/project/precision-medicine-toolbox/). An easy installation of the latest version is possible with “pip install precision-medicine-toolbox” command. At the time of submission of this work, \textit{precision-medicine-toolbox} is at 0.0 release.
The project has the following structure:
\begin{itemize}
    \item README.MD: file with the project overview,
    \item Requirements.txt: file with the list of the packages to be installed,
    \item LICENSE: statement of the license applicable to the project's software and manuscripts,
    \item .gitignore: specification of the files, intentionally untracked by Git,
    \item .readthedocs.yaml: Read the Docs configuration file, \item Mkdocs.yml: Mkdocs configuration file,
    \item Setup.cfg and setup.py: configuration files for PyPi package,
    \item Data: folder with the raw data for the examples as well as generated files,
    \item Docs: folder with the documentation files,
    \item Examples: folder with the examples:
    \begin{itemize}
        \item Imaging\textunderscore module.ipynb: tutorial illustrating functionality for the imaging datasets, 
        \item Features\textunderscore module.ipynb: tutorial illustrating functionality for the features datasets, 
    \end{itemize}
    \item Pmtool: folder with the toolbox scripts:
    \begin{itemize}
        \item \textunderscore \textunderscore init\textunderscore \textunderscore .py: initialization file,
        \item data\textunderscore set.py: script defining the base class for imaging datasets,
        \item tool\textunderscore box.py: script defining the inheriting class for imaging datasets methods,
        \item features\textunderscore set.py: script defining the base class for features datasets,
        \item analysis\textunderscore box.py: script defining the inheriting class for features datasets methods.
    \end{itemize}
\end{itemize}

The next sections shortly summarize the examples that cover the current functionalities of the \textit{precision-medicine-toolbox}.

\subsubsection{Imaging module}
The example ‘Imaging module’ illustrates how to explore the imaging parameters retrieved from the DICOM tags, perform data quality check, convert DICOM slices to volume format, perform image basic pre-processing, check ROI segmentation, and extract the radiomic features. 
At first, the \textit{ToolBox} class needs to be initialized with the user-defined parameters, such as the path to the dataset, data format, mask availability, mask file names, and image file names. After the \textit{ToolBox} object is created, the corresponding methods can be called. In the example, to speed up the process, we read only one mask per patient.
To get an insight of the data and plan the following pre-processing routines, we perform the exploration of the dataset by collecting its imaging metadata. After the initialization of the \textit{ToolBox} object, the \textit{get\textunderscore dataset\textunderscore description} method is implemented. The outcome is stored in the \textit{dataset\textunderscore description} DataFrame. If we call this method with the default parameters, we get modality, slice thickness, pixel spacing, date, and manufacturer for every DICOM file. After calling this method with the indication of the imaging modality (CT), the collected information contains patient name, CT convolution kernel, slice thickness, pixel spacing, kilovoltage peak, exposure, X-Ray tube current, and series date. For a better understanding of the data we are dealing with, we are using Python Matplotlib functionality to plot the distributions of these parameters. The results are presented in Figure 3.
\break\break\break
\includegraphics[scale=0.41]{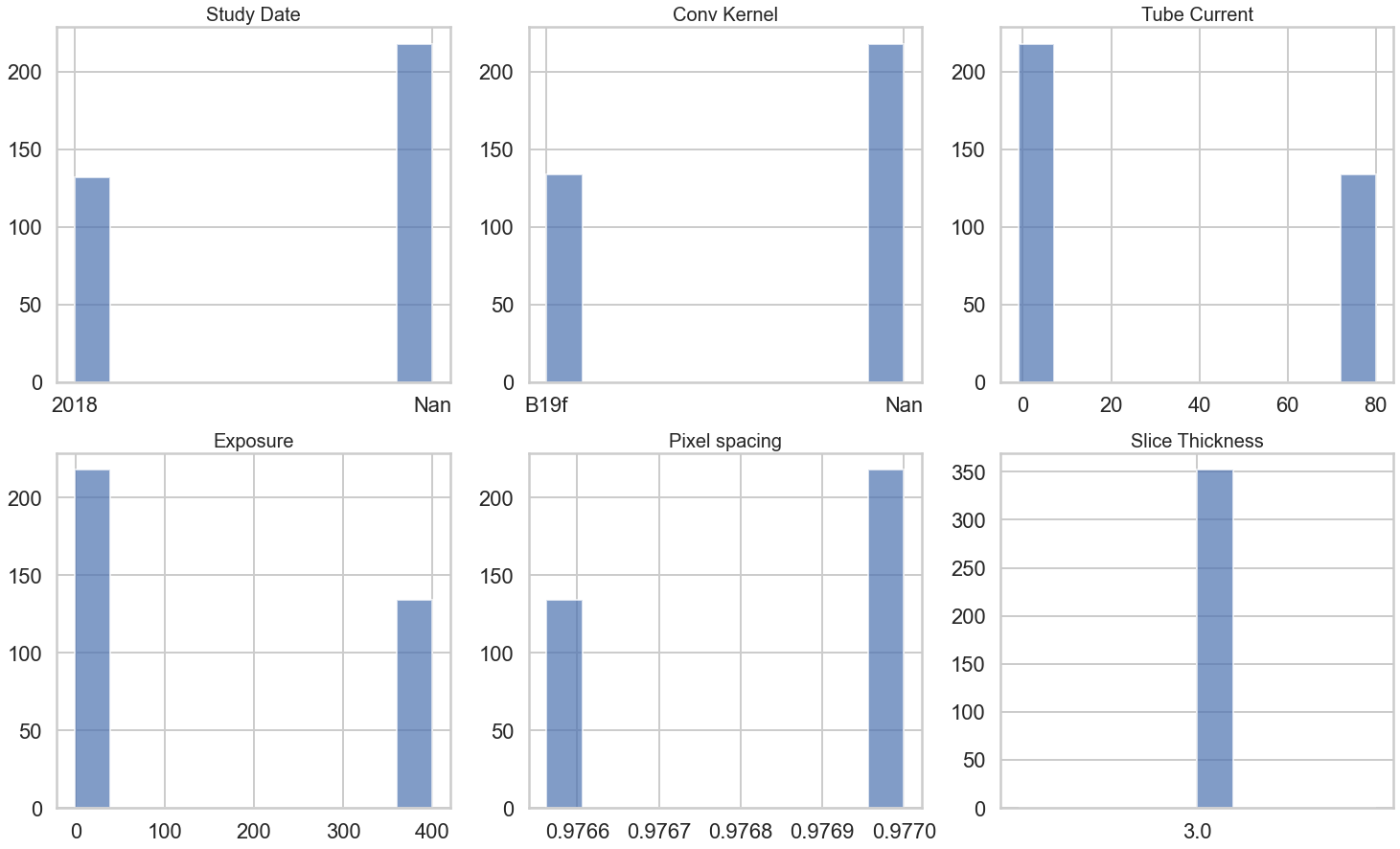}
\centerline{ Figure 3 - Distributions of some of the CT imaging parameters in Lung1 data set.}
\break\break
The \textit{get\textunderscore quality\textunderscore checks} method allows to perform a simple quality check of the data and possibly detect irrelevant scans. These might be scans of wrong imaging modality, with wrong imaging projections, with non-consistent (missing/overlapping) slices, with insufficient amount of slices, with slice thickness inconsistent or out of the defined range, with pixel spacing out of range, with unknown or unacceptable convolution kernel, with wrong axial plane resolution, with missing slope/intercept tags. To perform this check, the target scanning parameters are to be passed to the function. While removing some of the input parameters, the corresponding checks are disabled. For each patient, the output DataFrame contains the following flags: '1' - check passed, '0' - check failed.
The \textit{convert\textunderscore to\textunderscore nrrd} method is converting the DICOM data into volumetric NRRD format and saves it into the created folder (‘.../data/converted\textunderscore nrrd/’). Currently supported modalities are: CT, MRI, PET,  RTSTRUCT. In the example, we performed conversion of the DICOM dataset with CT and corresponding RTSTRUCTs containing GTV contours.
Image basic pre-processing is performed by the \textit{pre\textunderscore process} method. According to IBSI recommendations, radiomic analysis should be performed for the raw images, except for the modalities, represented in arbitrary units (e.g., MRI, ultrasound). For these modalities, Z-scoring is recommended. Nevertheless, some image pre-processing can be performed to keep the same data shape within the dataset, decrease diversity of the data, or harmonize the images from different datasets. The following functionality is available in the \textit{pre\textunderscore process} method: N4 bias field correction (\cite{Tustison2010-gf}), intensity rescaling and normalization, histogram matching and histogram equalization, intensity resampling, image reshaping. The pre-processing step is not executed, if the corresponding parameter is not passed to the method. It is possible to visualise every pre-processing step for every patient and print out the processing parameters and basic intensity statistics for input and output scans.
To perform a sanity check of the converted images and masks and their co-alignment, we initialize a \textit{ToolBox} class for the newly converted NRRD dataset. Then we call the \textit{get\textunderscore jpegs} method, which saves the converted JPEG slices into the …/data/‘images\textunderscore quick\textunderscore check/’ folder. The example of the output is presented in Figure 4.
\break\break\break
\includegraphics[scale=0.55]{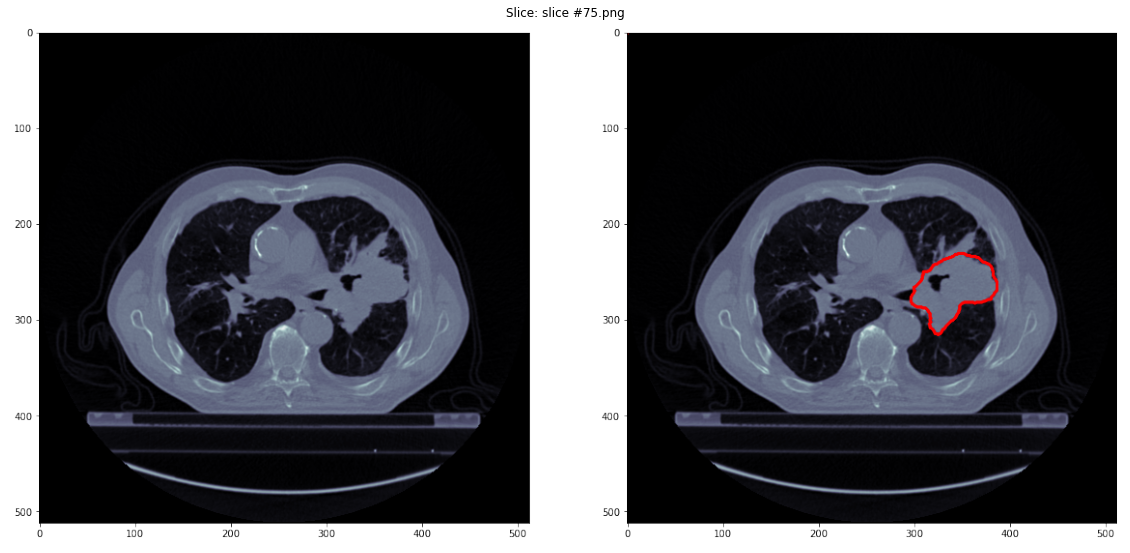}
\centerline{Figure 4 - Example of the quick segmentation check.}
\break\break
To extract the PyRadiomics features, a \textit{ToolBox} class for the newly converted NRRD dataset needs to be initialized. Then the features are extracted with the \textit{extract\textunderscore features} method. The extraction parameters are imported from the example\textunderscore ct\textunderscore parameters.yaml file. The parameter file is required by the PyRadiomics package and contains information about the preferred image types, features to extract, resampling and discretization settings. In the example we use the parameters suggested for the CT features extraction provided by the PyRadiomics repository. The \textit{extract\textunderscore features} method returns a Pandas DataFrame with radiomics features which can be further exported into the Excel file, CSV or any other format supported by Pandas.

\subsubsection{Features module}

The example ‘Features module’ illustrates how to visualize features distribution in classes, plot the feature correlation matrix, check Mann-Whitney U-test p-values, plot univariate ROC and calculate AUC for each feature, perform volumetric analysis, and save all the scores.
The tutorial is using the radiomics features, extracted from the Lung1 dataset, and the clinical data file, provided with the dataset. Using the clinical data, we generated three binary outcomes of 1-, 1.5-, and 2-years survival. In the tutorial, we present two cases: binary class dataset and multi-class dataset.
The \textit{AnalysisBox} class is calling a \textit{FeaturesSet} initialization with the user-defined parameters, such as paths to the tabular data with the features and outcomes, a list of the features to be included or excluded, names of the patient and outcome columns, and a list of the patients to be excluded. The dataset estimated parameters are the available class labels and dataset balance in terms of the outcome values.
For the binary class dataset, we declared 1yearsurvival as an outcome column. After \textit{AnalysisBox} object initialization, we get the class labels (‘0’ and ‘1’) and class balance (0.42 and 0.58).
After using the \textit{handle\textunderscore nan} method for the patients, there were no changes in the dataset, which means we did not have any missing values.
After calling the \textit{plot\textunderscore distribution} method, for each feature, the value distributions were plotted as bin histograms. The result is presented in Figure 5A. The class affiliation is highlighted with a color. The class label is presented on the plot.
After calling the \textit{plot\textunderscore correlation\textunderscore matrix} method, the mutual feature correlation coefficient (Spearman's) matrix is visualized. The result is presented in Figure 5B. The values are the absolute values. Colorbar is presented on the right side of the matrix.
After calling the \textit{plot\textunderscore MW\textunderscore p} method, Mann-Whitney (with Bonferroni correction) p-values for binary classes test are visualized as a barplot. The result is presented in Figure 5C. The p-value scale is logarithmic. If the p-value for some feature is below the set significance level (alpha=0.05), the corresponding bar is highlighted with a yellow color, whereas the other bars are purple. 
After calling the \textit{plot\textunderscore univariate\textunderscore roc} method, the univariate ROC curves are visualized for all the features. The ROC AUC scores are reported as well. The result is presented in Figure 5D. If the ROC AUC score is exceeding the set threshold (auc\textunderscore threshold=0.70), the curve is highlighted with the purple color. Otherwise, it is yellow.
After calling the \textit{volume\textunderscore analysis} method with sending there a volumetric feature name ('original\textunderscore shape\textunderscore VoxelVolume'), a volume precision-recall curve is visualized (with AUC calculated) as well as a barplot with volume Spearmen’s correlation coefficient absolute values with all the features. The resulting plots are presented in Figures 5E and 5F. If the correlation coefficient exceeds a threshold value (corr\textunderscore threshold=0.75), the bar is highlighted with the purple color. Otherwise, it is yellow.
After calling the \textit{calculate\textunderscore basic\textunderscore stats} method, the basic statistics are calculated for all the features. As the dataset has two classes, Mann-Whitney test p-values and univariate ROC AUC scores are calculated. We also define the feature, which is representing the volume ('original\textunderscore shape\textunderscore VoxelVolume'), thus Spearman’s correlation coefficient with volume is calculated. The results are saved into the ‘extracted\textunderscore features\textunderscore full\textunderscore basic\textunderscore stats.xlsx’ file, which belongs to the same directory as the features file.
\break\break\break
\includegraphics[scale=0.4]{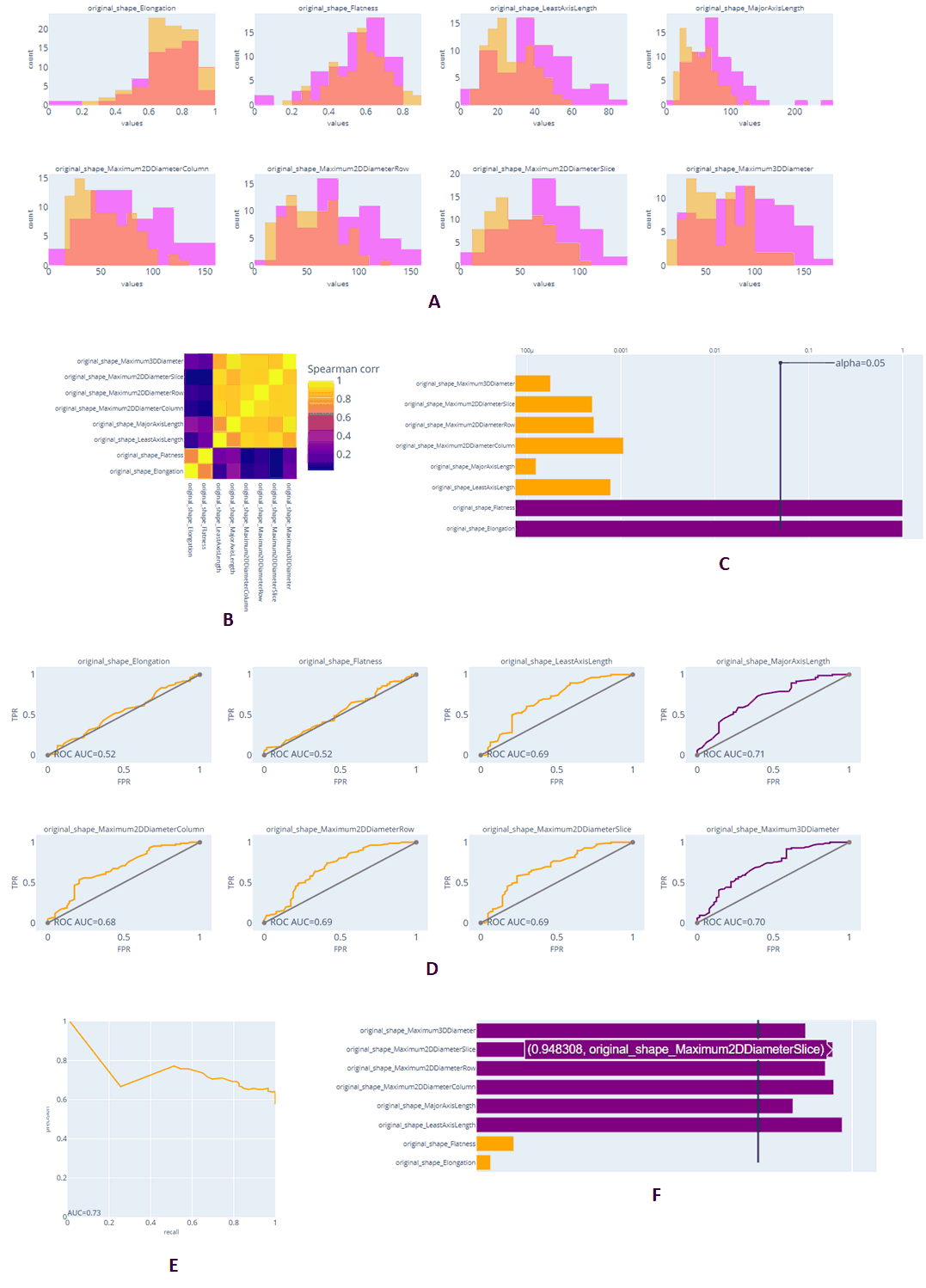}
\begin{tabular}{c}
Figure 5 - Feature analysis plots for binary outcomes for eight features: \\ A - feature value distributions in binary classes, B - Spearman’s correlation matrix between features,  \\ C - Mann-Whitney test (Bonferroni corrected) p-values, D - univariate ROC curves for binary classification, \\ E - volume based precision-recall curve, F - features Spearman's correlation with volume.
\end{tabular}
\break\break
The next part of the tutorial is devoted to multi-class analysis. The AnalysisBox is initialized in the same way, but the outcome column is changed to ‘Overall.Stage’. The available class labels are ‘I’, ‘II’, ‘IIIa’, ‘IIIb’, and the empty value. The class balance is 0.24, 0.09, 0.23, 0.42, and 0.01, respectively. While implementing the \textit{handle\textunderscore nan} method at the patient's level, one patient with an unknown outcome is dropped. Therefore, after the re-initialization of the class object, we have 148 patients with ‘I’, ‘II’, ‘IIIa’, and ‘IIIb’ labels. The class proportions are 0.24, 0.09, 0.23, and 0.43, respectively.
The \textit{plot\textunderscore distribution} method works for all the presented classes as well as for the selected classes. The result is presented in Figure 6.
\break\break\break
\includegraphics[scale=0.85]{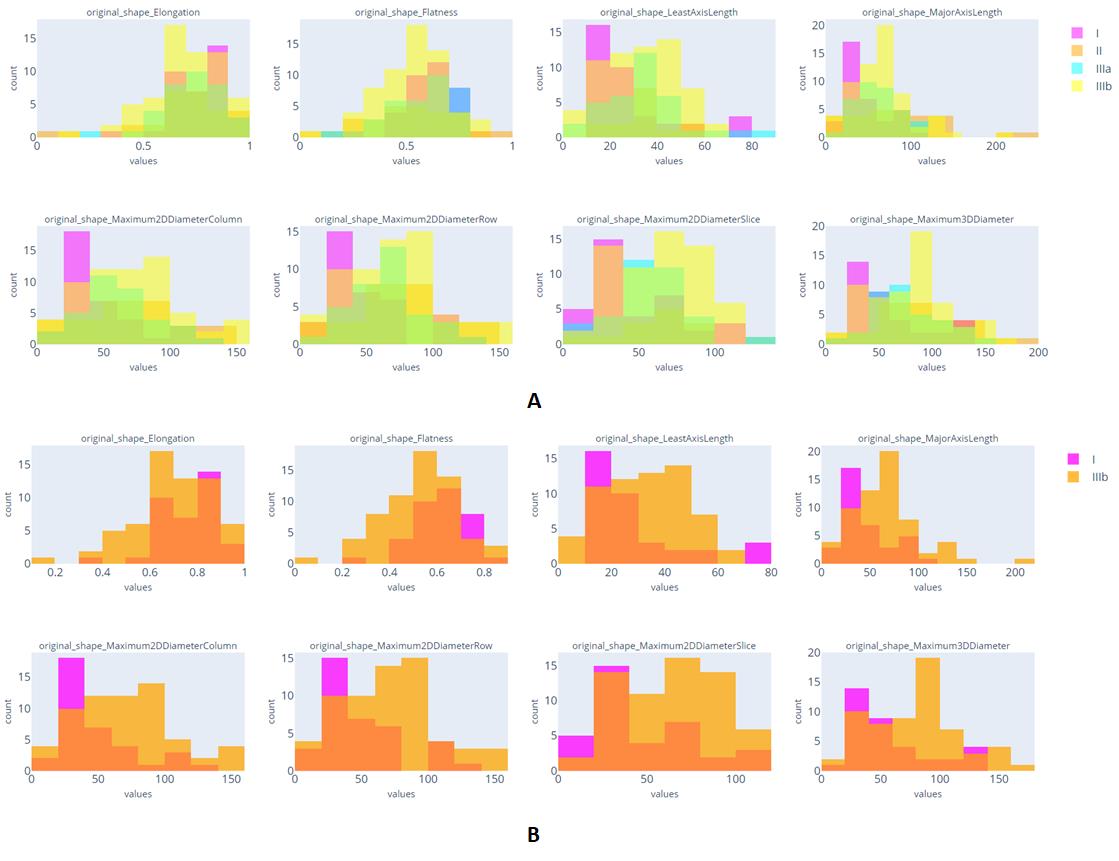}
\begin{tabular}{c}
Figure 6 - Feature value distributions in multiple classes: A - for all the presented classes, B - for the selected \\ classes I and IIIb.
\end{tabular}
\break\break
The \textit{plot\textunderscore MW\textunderscore p} and \textit{plot\textunderscore univariate\textunderscore roc} methods are not supported for the multi-class data, but they can still be implemented for the observations from any of the selected two classes. The other methods are working in the same way, as for binary classes. The \textit{calculate\textunderscore basic\textunderscore stats} method does not calculate Mann-Whitney test p-values and univariate ROC AUC scores.

\subsection{Online documentation and tutorials}

The documentation (http://precision-medicine-toolbox.readthedocs.io/) is built with Mkdocs (https://www.mkdocs.org/) and hosted on the Read the Docs platform (http://readthedocs.io). Code quality is reviewed with CodeFactor (http://codefactor.io). The API specifications for all the classes and methods are generated automatically from the source code annotations with Mkdocstrings (https://mkdocstrings.github.io/). This enables keeping documentation up to date with the latest developments of the package. The documentation also contains links to tutorials with examples generated from Jupyter notebooks. These notebooks are included in the \textit{precision-medicine-toolbox} package (https://github.com/primakov/precision-medicine-toolbox/tree/master/examples) and are available for any user.

\section{Discussion}
This paper introduced the open-source \textit{precision-medicine-toolbox} for imaging data preparation and exploratory analysis. It aims to address the data preparation and exploration problem, bridge the gap between the currently existing packages and improve the reproducibility of quantitative medical imaging research.

The functionality of the toolbox aims to meet some challenges that are specific to the radiomics field. One of these challenges is the lack of data and pipelines standardisation (\cite{Van_Timmeren2020-hx,Ibrahim2021-gf,Zwanenburg2020-ca}). Therefore, reproducibility is one of the key criterias for the radiomics studies. The \textit{precision-medicine-toolbox} has the functionality for the preliminary data check, including both investigation of the imaging parameters and features properties. This enables a rapid evaluation of the existing data, models, and studies. The other challenge is related to the large amount of the volume surrogate features. This means that many features are highly correlated with volume and do not add any value. In order to identify such features, volumetric analysis functions have been implemented in the \textit{precision-medicine-toolbox}.

The toolbox is mostly dedicated for the radiomics analysis, as it allows for handling of both raw imaging data and derivative features. Nevertheless, its modules can be used separately for other medical imaging research applications. Imaging module is applicable for the deep learning tasks to prepare the imaging data and get the information about its inhomogeneity. Features module can be used for any tabular data analysis, such as health records variables, or histology-derived features.

The \textit{precision-medicine-toolbox} was successfully utilised and tested during the development of multiple projects including DUNE.AI, automatic NSCLC segmentation on the CT (\cite{Primakov2021-cs}), repeatability of breast MRI radiomic features (\cite{Granzier2021-zl}), prognostic and predictive analysis of Glioblastoma MRI (\cite{Verduin2021-sc}), quantitative MRI biomarkers discovery in multiple sclerosis (\cite{Lavrova2021-tf}).The development of \textit{precision-medicine-toolbox} not only aims for democratisation of the machine learning and deep learning pipelines for the researchers without strong programming skills but also drives a programming community effort to improve this package and add its own variables and methods. Therefore, user contributions are very welcome.

\section{Conclusions}

The development of \textit{precision-medicine-toolbox} aims to lower the entry barrier for researchers who are starting to work in medical imaging and provide an open source solution for the researchers who already have their inhouse workflow of managing data to increase the reproducibility of the quantitative medical imaging research. We would also like to encourage the community to improve this open-source toolbox by contributing to it.

\section{Acknowledgements}

The authors would like to thank the Precision Medicine department colleagues for their support and feedback. We also would like to thank PyRadiomics authors for a reliable open-source tool for radiomic features extraction. And we would like to thank TCIA for the open-source Lung1 dataset we used to demonstrate our functionality.

\bibliographystyle{unsrtnat}
\bibliography{references}  %%% Uncomment this line and comment out the ``thebibliography'' section below to use the external .bib file (using bibtex) .
\end{document}